\RequirePackage[l2tabu]{nag} 
\documentclass[10pt]{article}

\usepackage[utf8]{inputenc}
\usepackage[T1]{fontenc}
\usepackage[top=1.2in, bottom=1.35in, left=1.4in, right=1.4in]{geometry}

\usepackage{lmodern}
\usepackage{microtype}
\microtypesetup{kerning=true,tracking=true,spacing=true,factor=1100,stretch=10,shrink=10}
\SetExtraKerning[unit=space]
    {encoding={*}, family={cmr}, series={*}, size={footnotesize,small,normalsize}}
    { 
     "28={ ,100}, 
     "29={100, } 
    }
\usepackage{bbm}

\usepackage{bm}
\usepackage[intlimits]{mathtools}
\usepackage{amssymb}
\usepackage{amsthm}

\usepackage[dvipsnames]{xcolor}
\usepackage[unicode=true,pdfusetitle, bookmarks=true,bookmarksnumbered=false,bookmarksopen=false, breaklinks=false,hidelinks,colorlinks=true,hypertexnames=false]{hyperref}
\hypersetup{colorlinks=true, linkcolor=Blue,citecolor=Magenta, urlcolor=Blue}

\usepackage[sorting=nyt,doi=false,eprint=false,url=false,isbn=false,maxbibnames=20,  backend=biber]{biblatex}
\usepackage{caption}
\usepackage{subcaption}
\usepackage{color}
\usepackage{graphicx}
\usepackage{epstopdf}

\theoremstyle{plain}

\usepackage[noabbrev]{cleveref}
\crefname{lem}{lemma}{lemmas}

\usepackage{soul}%
\usepackage{float}
\usepackage{booktabs}

\DeclarePairedDelimiter\abs{\lvert}{\rvert}
\DeclarePairedDelimiter\norm{\lVert}{\rVert}

\DeclareMathOperator{\expec}{\mathbb E}

\DeclareMathOperator{\corr}{Corr}
\newcommand{\prob}{\mathbb P}
\newcommand{\pdf}{f}

\DeclareMathOperator{\hist}{Hist}

\newcommand{\blank}{\,\cdot\,}

\newcommand{\e}{\mathrm{e}}

\usepackage{enumerate}

\usepackage{tabularx} 
\newcolumntype{C}{>{\centering\arraybackslash}X} 

\begin{document}

\title{\bfseries Heavy-tailed response of structural systems subjected to stochastic excitation containing extreme forcing events}

\author{Han Kyul Joo, Mustafa A. Mohamad,   Themistoklis P.  Sapsis\thanks{Corresponding author: \href{mailto:sapsis@mit.edu}{sapsis@mit.edu},
Tel: (617) 324-7508, Fax: (617) 253-8689}}
\date{Department of Mechanical Engineering,
    \\ Massachusetts Institute of Technology, \\
    77 Massachusetts Ave., Cambridge, MA 02139 \\ \medskip \today}
\maketitle
\begin{abstract}
    We characterize  the complex, heavy-tailed probability distribution functions (pdf) describing the response and its local extrema for structural systems subjected to random forcing that includes  extreme events. Our approach is based on the recent probabilistic decomposition-synthesis technique in~\cite{mohamad2016b}, where we decouple rare events regimes from the background fluctuations. The result of the analysis has the form of a semi-analytical approximation formula for  the pdf of the response (displacement, velocity, and acceleration) and the pdf of the local extrema. For special limiting cases (lightly damped or heavily damped systems) our analysis provides  fully analytical approximations. We also demonstrate how the method can be applied to high dimensional structural systems through a two-degrees-of-freedom structural  system undergoing rare events due to intermittent forcing. The derived formulas can be evaluated with very small computational cost and are shown to accurately capture the complicated heavy-tailed and asymmetrical features in the probability distribution many standard deviations away from the mean, through comparisons with expensive Monte-Carlo simulations. 
\end{abstract}


\section{Introduction}

\begin{figure}\centering
    \includegraphics[width=0.85\textwidth]{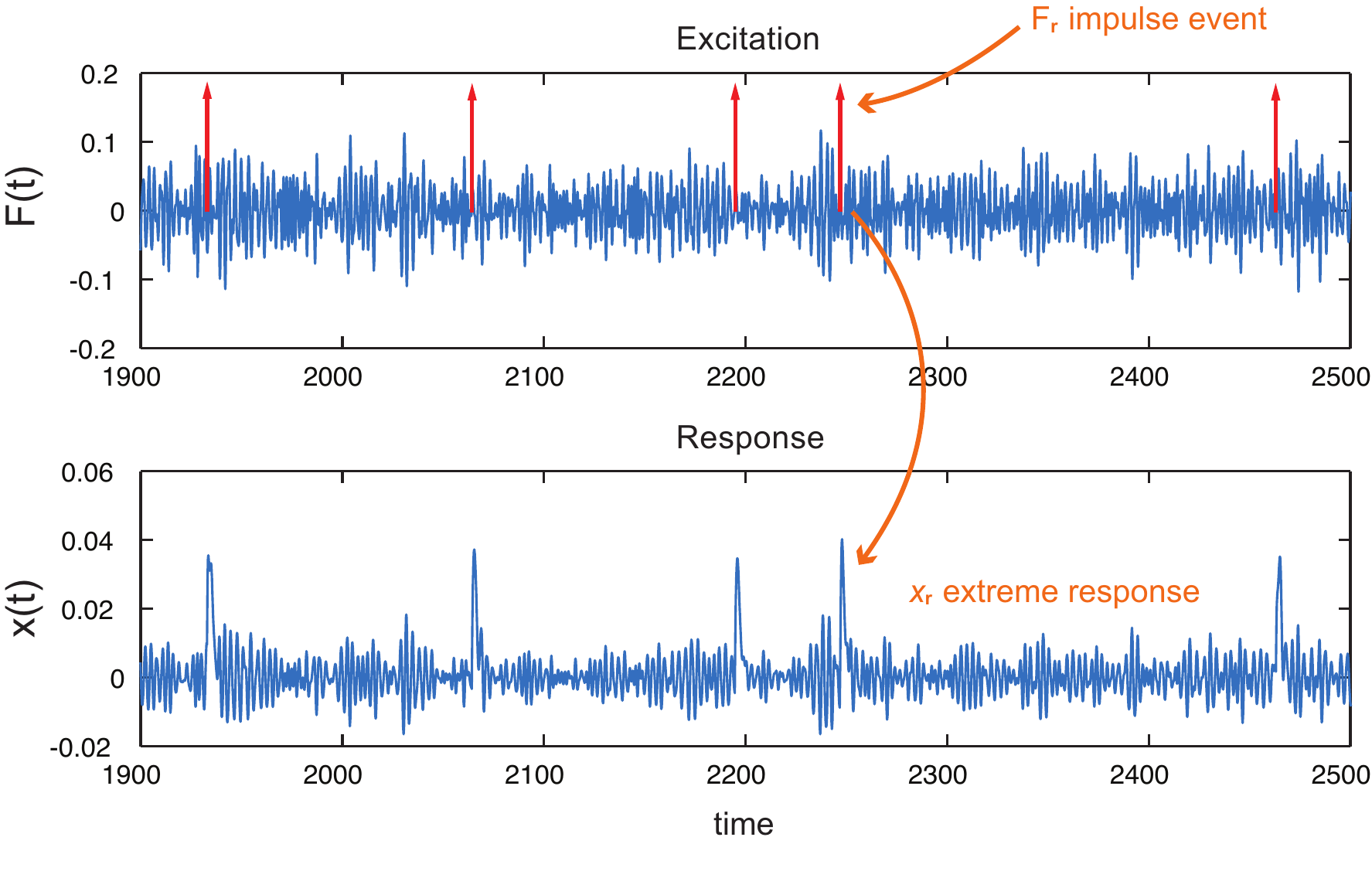}        
    \caption{(Top) Background stochastic excitation including impulsive
loads in (red) upward arrows. (Bottom) System  response displacement.}
\label{figure1}
\end{figure}

A large class of physical systems in engineering and science can be   modeled  by stochastic differential equations. For many of these systems, the  dominant source of uncertainty is due to the forcing  which can be described  by a stochastic process.  Applications include ocean engineering systems excited by water waves (such as ship motions in large waves ~\cite{mohamad2016a,belenky07,cousinsSapsis2015_JFM,cousins_sapsis} or high speed crafts subjected to rough seas ~\cite{Riley2011, Riley2012}) and rare events in structural systems (such as  beam buckling~\cite{Abou-Rayan1993,Lin_Cai95}, vibrations due to  earthquakes~\cite{Lin63, Lin98} and wind loads \cite{lin96, gioffre12}).  For all of these cases it is common that hidden in the otherwise predictable magnitude of the  fluctuations  are   extreme events, i.e. abnormally large magnitude forces which lead to rare responses in the dynamics of the system (\cref{figure1}). Clearly, these events must be adequately taken into account for the effective quantification of the reliability properties of the system.   In this work, we develop an efficient method to fully describe  the  probabilistic response of linear structural systems under general time-correlated random  excitations containing rare and extreme events.

Systems with forcing having these characteristics pose significant challenges for traditional
uncertainty quantification schemes. While there is a large class of methods  that can accurately resolve the statistics associated with random excitations  (e.g. the Fokker-Planck equation~\cite{Soong93,sobczyk01} for systems excited by white-noise and the joint response-excitation method~\cite{Sapsis08, venturi_sapsis, Joo16, athan_2016_prsa} for arbitrary stochastic excitation) these have important limitations for high dimensional systems. In addition, even for low-dimensional systems determining the part of the probability density function (pdf)  associated with extreme events poses important numerical challenges. On the other hand, Gaussian closure schemes and moment equation or cumulant closure methods~\cite{Beran68, wu84,} either cannot ``see'' the rare events completely or they are very expensive and require the solution of an inverse moment problem in order to determine the pdf of interest~\cite{Athanas_Gavrilia_Housdorff}. Similarly, approaches relying on polynomial-chaos expansions~\cite{Xiu_Karniadakis02, Xiu_Karniadakis03} have been shown to have important limitations for systems with intermittent responses~\cite{majda_branicki_DCDS}.

Another popular approach for the study of rare event statistics in  systems under intermittent forcing is to   represent   extreme events in the forcing  as identically distributed independent impulses arriving at random times. The generalized Fokker-Planck equation or Kolmogorov-Feller (KF) equation is the governing equation that solves for  the evolution of the response pdf under Poisson noise~\cite{sobczyk01}. However,   exact analytical solutions are available only for a limited number of special cases~\cite{Vasta95}. Although alternative methods such as the  path integral method~\cite{koyluoglu95, Iwankiewicz00, Barone08} and the stochastic averaging method~\cite{Zhu88, Zeng11} may be applied, solving the FP or KF equations is often very expensive ~\cite{Masud_bergman05,di14} even for very low dimensional systems.

Here, we consider the problem of quantification of the response pdf and the pdf associated with local extrema of  linear systems subjected to stochastic forcing containing extreme events based on  the recently formulated probabilistic-decomposition  synthesis (PDS) method~\cite{mohamad2016b,mohamad2015}. The approach relies on the decomposition of the statistics into a `non-extreme core', typically Gaussian, and a heavy-tailed component. This decomposition is in full correspondence with a partition of the phase space into a `stable' region where we do not have rare events and a region where non-linear instabilities or external forcing lead to rare transitions with high probability. We quantify the statistics in the stable region using a Gaussian approximation approach, while the non-Gaussian distribution associated with the intermittently unstable regions of phase space is performed taking into account the  non-trivial character of the dynamics (either because of instabilities or external forcing). The probabilistic information in the two domains is analytically synthesized through a total probability argument.  

 We begin with the simplest case of a linear, single-degree-of-freedom (SDOF) system   and then formulate the method  for multi-degree-of-freedom
systems. \emph{The main result of our work  is the derivation of analytic/semi-analytic approximation
formulas for the response pdf  and the pdf of the local extrema of intermittently forced systems that
can accurately characterize the statistics many standard deviations away from the mean}.  Although the systems considered in this work are linear the method is directly applicable for nonlinear structural systems as well. This
approach circumvents   the challenges that rare events pose for traditional
uncertainty quantification schemes, in particular  the computational burden associated when dealing
with rare events in systems. We emphasize the statistical accuracy and the computational efficiency
of the presented  approach, which we rigorously demonstrate through extensive comparisons with
direct  Monte-Carlo simulations. In brief, the principal contributions of this paper are:
\begin{itemize}
    \item Analytical (under certain conditions) and semi-analytical (under no restrictions) pdf expressions for the response displacement, velocity and acceleration for single-degree-of-freedom systems under intermittent forcing.
    \item Semi-analytical pdf expressions for the value and the local extrema of the displacement, velocity and acceleration for multi-degree-of-freedom systems under intermittent forcing.
\end{itemize}

The paper is structured as follows. In~\cref{sec:formu}   we provide a  general formulation of the probabilistic decomposition-synthesis method for the  case of structural systems under intermittent forcing. Next, in~\cref{sec:sdof_ana}, we apply the developed method analytically, which is  possible for two limiting cases:  underdamped systems with  $\zeta\ll1$ or   overdamped with  $\zeta\gg1$, where $\zeta$ is the damping ratio. The system we consider is  excited by a forcing term consisting of a background time-correlated stochastic process  superimposed with   a random impulse train (describing the rare and extreme component).  We give a detailed derivation of the   response pdf of the system  (displacement, velocity and acceleration)  and compare the results   with expensive  Monte-Carlo  simulations.  In~\cref{sec:sdof_semi}, we slightly modify the developed formulation to derive a semi-analytical scheme considering the same linear system but without any restriction on the damping ratio $\zeta$, demonstrating global applicability of our approach.  In~\cref{sec:2dof_semi}, we  demonstrate applicability of our method for multiple-degree-of-freedom systems and in section 6 we present results for the local extremes of the response. Finally, we offer   concluding remarks in~\cref{sec:conc}.

\section{The probabilistic decomposition-synthesis method for intermittently forced structural systems}\label{sec:formu}

We provide with a brief presentation of the recently developed probabilistic decomposition-synthesis (PDS) method adapted for the    case of  intermittently forced linear structural systems \cite{mohamad2016b}. 
We consider the following vibrational system,
\begin{equation}\label{eq:original_sys}
    M \ddot{\mathbf{x}}(t) + D\dot{\mathbf{x}}(t) + K{\mathbf{x}}(t) = \mathbf F(t), \quad \mathbf x(t) \in \mathbb R^n,
\end{equation}
where $M$ is a mass matrix, $D$ is the   damping matrix,  and $K$ is the stiffness matrix. We assume $\mathbf F(t)$ is a stochastic forcing with intermittent characteristics that can be expressed as
\begin{equation}
    \mathbf F(t) = \mathbf F_b(t) + \mathbf F_r(t).
\end{equation}
The forcing consists of a background component $\mathbf F_b$ of characteristic magnitude  $\sigma_b$ and a \emph{rare and extreme} component $\mathbf  F_r$ with   magnitude  $\sigma_r \gg \sigma_b.$  The   components $\mathbf F_b$ and $\mathbf  F_r$ may both be (weakly) stationary stochastic processes, while the sum of the two processes can be, in general,  non-stationary.
This can be seen if we directly consider the sum of two (weakly) stationary processes $x_1$ and $x_2$, with time correlation functions $ \corr_{x_1}(\tau)$ and $ \corr_{x_2}(\tau)$, respectively. Then for the sum $z = x_1  + x_2 $ we have
    \[
        \corr_z(t,\tau) =  \corr_{x_1}(\tau) +   \expec[x_1 (t) x_2(t+\tau)] +  \expec[x_1(t+\tau) x_2(t)] + \corr_{x_2}(\tau).
    \]
Therefore, the process   $z$ is stationary if and only if    the cross-covariance terms  $\expec[x_1 (t) x_2(t+\tau)]$ and $\expec[x_1 (t+\tau) x_2(t)]$ are functions of $\tau$ only or they are zero (i.e. $x_1$ and $x_2$ are not correlated).

For the case where the excitation is given in terms of realizations, i.e. time-series, one can first separate the extreme events from the stationary background by applying time-frequency  analysis methods (e.g. wavelets~\cite{wavelet_sri}). Then the stationary background can be approximated with a Gaussian stationary stochastic process (with properly tuned covariance function) while the rare event component can be represented with a Poisson process with properly chosen parameters that represent the characteristics of the extreme forcing events (frequency and magnitude).

\begin{figure}
    \centering
    \includegraphics[width=0.85\textwidth]{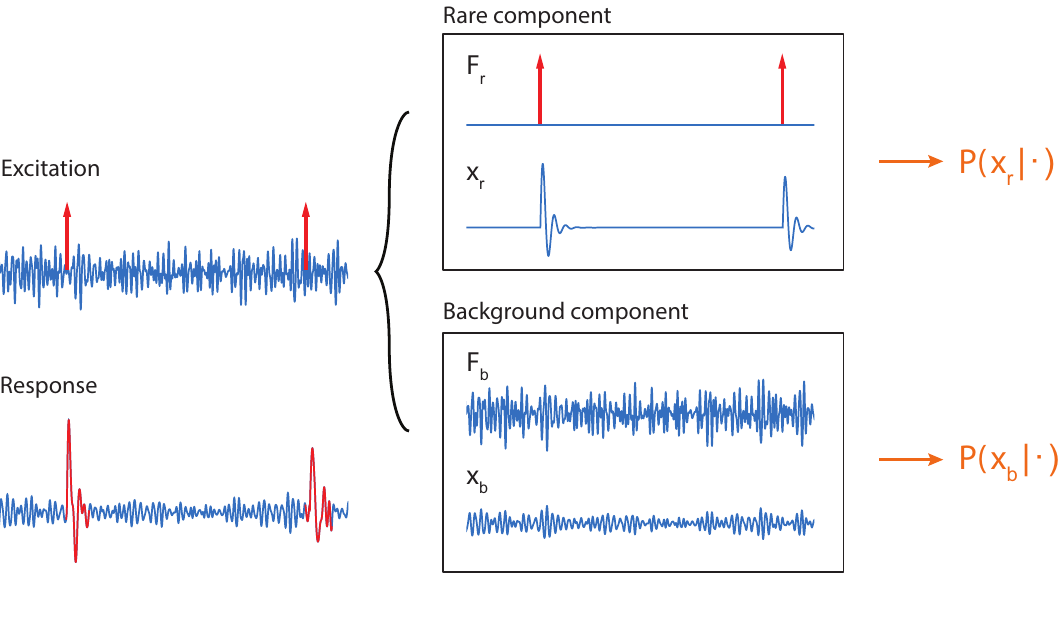}
    \caption{Schematic representation of the PDS method for an intermittently forced system.}
\label{figure2}
\end{figure}

To apply the PDS method we also decompose the response into  two terms
\begin{equation}
    \mathbf x(t) = \mathbf x_b(t) + \mathbf x_r(t),
\end{equation}
where $\mathbf{x}_b$ accounts for the background state (non-extreme) and $\mathbf{x}_r$ captures
extreme responses (due to the intermittent forcing) -- see~\cref{figure2}. More precisely   
$\mathbf x_r$ is the system response  under two conditions:  (1) the forcing is given by   $\mathbf  F =  \mathbf F_r$    (i.e. we have an impulse) and   (2)  the norm of the response is greater than the  background response fluctuations according to a given criterion, e.g. $\left\Vert \mathbf{x} \right\Vert >\gamma$. However, as we will see in the following sections other criteria may be used. These rare transitions  occur when we have an impulse and they also include a phase that  relaxes the system back to the background state $\mathbf x_b$. The background component $\mathbf x_b$ corresponds to system     response without  rare  events $\mathbf x_b = \mathbf x - \mathbf x_r$, and in this regime  the system  is  primarily governed by the background forcing term $\mathbf F_b.$ 

We require that rare events are statistically independent from each other. In the  generic formulation of the PDS we also need to assume that rare events have negligible effects on the background state $\mathbf{x}_b$  but here this assumption is not necessary due to the linear character of the examples considered. However, in order to apply the method for general nonlinear structural systems we need to have this condition satisfied.
We also need the dynamics to be ergodic while the statistics we aim to approximate refer to long time averages \cite{mohamad2016b}.

Next, we focus on the statistical characteristics of an individual mode $u(t)\in \mathbb R$ of the original system in~\cref{eq:original_sys}. The first step of the PDS method is to quantify the conditional statistics of the rare event regime. When the system enters  the rare event response  at $t=t_0$ we will have an arbitrary background state $u_{b}$ as an initial condition at $t_0$ and the problem will be formulated as:
\begin{equation}  \label{eq:rare_symp}
    \ddot u_r(t) + \lambda  \dot u_r(t)   + k  u_r(t)  = F_r(t), \text{ }  \quad\text{ with  $u_{r}(t_{0})=u_b$  and $F  = F_r $  for $t> t_0$}.
\end{equation}
 Under  the assumption of independent rare events we can use~\cref{eq:rare_symp} as a basis to derive analytical or numerical estimates for the statistical response during the rare event regime.

The background component, on the other hand, can be studied through the equation,
\begin{equation}
    M \ddot{\mathbf{x}}_b(t) + D\dot{\mathbf{x}_b}(t) + K{\mathbf{x}_b}(t) = \mathbf F_b(t).
\end{equation}
Because of the non-intermittent character of the response in this regime, it is sufficient to obtain the low-order statistics of this system. In the context of vibrations it is reasonable to assume that $F_b(t)$ follows a Gaussian distribution in which case the problem is straightforward. Consequently, this step provides us with the statistical steady state probability distribution for the mode of interest under the condition that the dynamics `live' in the stochastic background.

Finally, after the  analysis of the two regimes is completed we can synthesize the results  through a   total probability argument 
\begin{equation}
    \pdf(q) = \underbrace{\pdf(q \mid  \norm{u} > \gamma, F =  F_r)}_{\text{rare events}} \prob_r + \underbrace{\pdf(q \mid F =  F_b)}_{\text{background}}  (1-\prob_r),
\end{equation}
where $q$ may be any function of interest involving the response. In the last equation, $\prob_r$ denotes the overall rare event  probability. This  is defined as
the probability of the response exceeding a threshold $\gamma$ because of a rare event in the excitation: \begin{equation}\label{eq:rareprobdef}
    \prob_r \equiv \prob(\norm{u} > \gamma, F = F_r) = \frac{1}{T} \int_{t\in T}  \mathbbm 1(\norm{u}>\gamma, F = F_r) \, dt,
\end{equation}
where $\mathbbm 1(\blank)$ is the indicator function.  The rare event probability   measures the total duration of the rare events taking into account their frequency and duration.
The utility of the presented decomposition is its flexibility in capturing rare responses, since we can account for the rare event dynamics directly  and connect their statistical properties  directly to the   original  system response.

\subsection{Problem formulation for  linear SDOF systems}

To demonstrate the method we begin with a very simple example and we   consider a single-degree-of-freedom linear system
\begin{equation}
    \ddot{x} +\lambda \dot{x} +k x = F(t),
\end{equation}
where $k$ is the   stiffness, $\lambda$ is the damping,  $\zeta = \lambda/ 2\sqrt{k}$ is the damping ratio. For what follows we  adopt the standard definitions:   $\omega_n=\sqrt{k}$,
$\omega_o = \omega_n \sqrt{\zeta^2-1}$, and $\omega_d = \omega_n \sqrt{1-\zeta^2}$.    $F(t)$  is a stochastic forcing term with intermittent characteristics, which can be written as 
\begin{equation}
    F(t) = F_b(t) + F_r(t).
\end{equation}
Here $F_b$ is  the background forcing component   that has a   characteristic magnitude $\sigma_b$
and $F_r$  is  a   rare and large amplitude forcing  component that has a  characteristic magnitude
$\sigma_r,$ which is much larger than the magnitude of the background forcing,  $\sigma_r \gg
\sigma_b$. Despite the  simplicity of the system,  its response  may feature a significantly  
complicated statistical structure with heavy-tailed characteristics.

For concreteness, we  consider a   prototype   system   motivated from ocean  engineering applications, modeling base excitation of a structural mode:
\begin{equation}\label{eq:prototype_eg1}
    \ddot{x} +\lambda \dot{x} +k x = \ddot{h}(t) + \sum^{N(t)}_{i=1}  \alpha_i\, \delta(t-\tau_i),
    \quad 0 < t \leq T.
\end{equation}
Here  $h(t)$ denotes the zero-mean background   base motion term (having opposite sign from $x$)   with a   Pierson-Moskowitz spectrum:
\begin{equation}\label{eq:pmspectrum}
    S_{hh}(\omega) = q \frac{1}{\omega^5}\exp\biggl(-\frac{1}{\omega^4}\biggr),
\end{equation}
where $q$ controls the magnitude of the forcing. 

The second forcing term in~\cref{eq:prototype_eg1} describes rare and extreme events. In particular, we assume this component  is  a random impulse train ($\delta(\blank)$ is a  unit impulse), where $N(t)$ is a Poisson counting process that represents the number of impulses that arrive in the time interval $0 < t \leq T$, $\alpha$ is the  impulse mean  magnitude  (characterizing the rare event magnitude $\sigma_r$), which we assume is normally distributed with mean $\mu_\alpha,$  variance $\sigma_\alpha^2$ and independent from the state of the system. In addition, the   arrival rate is constant and given by  $\nu_\alpha$ (or by  the    mean arrival time $T_\alpha ={1}/{\nu_\alpha}$ so that  impulse arrival times are exponentially distributed $ \tau \sim e^{T_\alpha}$).

We take the  impulse mean magnitude as  being $m $-times larger than the standard deviation of the  excitation velocity $\dot h(t)$:
\begin{equation}
    \mu_\alpha = m \sigma_{\dot{h}}, \text{ with }   m  > 1,
    \label{imp_criterai}
\end{equation}
where $\sigma_{\dot{h}}$ is the standard deviation of $\dot h(t)$.  This prototype system  is widely applicable to numerous applications, including   structures under wind   excitations, systems
under seismic excitations, and vibrations of high-speed crafts and road vehicles~\cite{sobczyk01,Soong93, Riley2011}.

\section{Analytical pdf of SDOF systems for limiting cases}\label{sec:sdof_ana}

In this section, we   apply the probabilistic decomposition-synthesis method for the
special cases $\zeta \ll 1$ and $\zeta \gg 1$  to  derive   analytical approximations for the  pdf of the displacement,  velocity and acceleration.  We   perform the analysis first for the response displacement  and by way of a minor modification obtain the pdf for the velocity and acceleration.

\subsection{Background response pdf }\label{sec:backgroundpdf}

Consider the statistical response of the   system to  the background forcing component,
\begin{equation}
    \ddot{x_b}+\lambda \dot{x_b} +k x_b= \ddot{h}(t),
\end{equation}
due to the Gaussian character of the statistics the response is fully characterized by it's
spectral response. The spectral density of the displacement, velocity and acceleration of this
system are given by,
\begin{equation}
    S_{x_bx_b}(\omega) =  \frac{\omega^4S_{hh}(\omega)}{\{k-\omega^2+\lambda(j\omega) \}^2} , \quad
    S_{\dot{x}_b\dot{x}_b}(\omega)  =\omega^2 S_{x_bx_b}(\omega), \quad
    S_{\ddot{x}_b\ddot{x}_b}(\omega)  = \omega^4 S_{x_bx_b}(\omega).
\end{equation}
Thus, we can obtain the variance of the response displacement, velocity and acceleration:
\begin{equation}\label{eq:backgroundvariance}
    \sigma_{x_b}^2 =  \int^\infty_0 S_{x_b x_b}(\omega)\,d\omega , \quad
    \sigma_{\dot{x}_b}^2 =  \int^\infty_0 S_{\dot{x}_b\dot{x}_b}(\omega) \,d\omega, \quad
    \sigma_{\ddot{x}_b}^2 =  \int^\infty_0 S_{\ddot{x}_b\ddot{x}_b}(\omega) \,d\omega.
\end{equation}
Moreover, the  envelopes are Rayleigh distributed~\cite{267314419860322}:
\begin{equation}\label{eq:env_background}
    u_b \sim \mathcal{R}(\sigma_{x_b}), \quad  \dot{u}_b \sim \mathcal{R}(\sigma_{\dot{x}_b}), \quad  \ddot{u}_b \sim \mathcal{R} (\sigma_{\ddot{x}_b}).
\end{equation}

\subsection{Analytical pdf for the underdamped case~\texorpdfstring{$\bm{\zeta\ll1}$}{zeta
<< 1}} 
Because of the underdamped character of the response for the case of ${\zeta\ll1}$, we focus on deriving the statistics of the local extrema. To this end, we will be presenting results for the statistics of the envelope of the response.
\subsubsection{Rare events response} 

To estimate the rare event response we take into account the non-zero background velocity of the system $\dot x_b$ at the moment of impact, as well as the magnitude of the impact, $\alpha$. The actual value of the response $x_b$ is considered negligible. For this case, taking into account ${\zeta\ll1}$, we have the envelopes of the response (displacement, velocity, acceleration) during the rare event given by (see~\cref{sec:raretransitions} for details),
\begin{equation}\label{eq:rare_enveloperesp}
    u_{r}(t) \simeq  \frac{\abs{ \dot x_b + \alpha}}{\omega_d} \e^{-\zeta \omega_n t} , \quad \dot u_{ r}(t) \simeq   \abs{\dot x_b + \alpha } \e^{-\zeta \omega_n t}, \quad
    {\ddot u}_{r}(t) \simeq  \omega_d \abs{\dot x_b + \alpha }  \e^{-\zeta \omega_n t}.
\end{equation}
In~\cref{eq:rare_enveloperesp} the two contributions $\dot x_b$ and $\alpha$ in the   term $\dot x_b + \alpha$ are both Gaussian distributed and independent and therefore their sum is also Gaussian distributed as:
\begin{equation}\label{eq:pdf_eta}
     \eta \equiv  \dot x_b + \alpha  \sim \mathcal N(\mu_\alpha,\,  \sigma_{\dot{x}_b}^2 + \sigma_{\alpha}^2).
\end{equation}
Therefore,  the distribution of the  quantity $\abs{\eta}$ is given by the following folded normal distribution:
\begin{equation}
    \pdf_{\abs{\eta}}(n) = \frac{1}{\sigma_{\abs{\eta}}\sqrt{2\pi}}\biggl\{\exp\biggl(-\frac{(n-\mu_\alpha )^2}{2\sigma_{\abs{\eta}}^2}\biggr)+\exp\biggl(-\frac{(n+\mu_\alpha )^2}{2\sigma_{\abs{\eta}}^2}\biggr)  \biggr\} ,\quad 0<n<\infty,
    \label{eq:pdf_eta_fold}
\end{equation}
where $\sigma_{\abs{\eta}} = \sqrt{\sigma_{\dot{x}_b}^2 + \sigma_{\alpha}^2}$.

\subsubsection{Rare event probability}\label{sec:rareeventtransitionprob}
Next, we compute the rare event  probability, which is the total duration of the rare events over a
time interval, defined in~\cref{eq:rareprobdef}. This will be done by employing an appropriate description for extreme events. One possible option is to set an absolute threshold $\gamma$. However, in the current context it is more convenient to set this threshold relative to the local maximum of the response. Specifically, the  time duration
$\tau_e$ a rare response takes to return back to  the background state will be given by the  duration  starting  from  the initial impulse event time ($ t_0 $)  to the point where the
response has decayed back to $\rho_c $ (or $100 \rho_c\%$) of its absolute maximum; here and throughout this manuscript we take
$\rho_c=0.1$. This is a value that we considered without any tuning. We emphasize that the derived approximation is not sensitive to the exact value of $\rho_c $ as long as this has been chosen within reasonable values. 

This means that $\tau_e$ is defined by
\begin{equation}
    u_r(\tau_{e}+t_{0})= \rho_c u_r(t_{0}),
\end{equation}
We solve the above \textit{using the derived envelopes} to obtain 
\begin{equation}\label{eq:tendformula}
    \tau_{e} =- \frac{1}{\zeta\omega_n}\log \rho_c.
\end{equation}
We note that due to the linear character of the system \emph{the typical  duration $\tau_e$ is  independent of the background state or the impact intensity}. 
With the obtained value for $\tau_{e}$ we compute the  probability of rare events  using the frequency $\nu_\alpha$ (equal to $1/T_\alpha$) :
\begin{equation}
    \prob_r = \nu_\alpha \tau_e = \tau_e/ T_\alpha.
\end{equation}
Note that based on our assumption that extreme events are rare enough to be statistical independent, the above probability is much smaller than one.

\subsubsection{Conditional pdf for rare events}\label{sec:rarepdf}

We now proceed with the derivation of the pdf in the rare event regime. Consider again the response
displacement during a rare event,
\begin{equation}
    u_r(t) \sim   \frac{ \abs{\eta}}{\omega_d} \e^{-\zeta \omega_n t'} ,
\end{equation}
here  $t'$ is a random variable uniformly distributed between the initial impulse event and the end time $\tau_e$ (\cref{eq:tendformula}) when the response has relaxed back to the background dynamics:
\begin{equation}
    t' \sim \text{Uniform}(0,\, \tau_e).
\end{equation}We condition the rare event distribution as follows,
\begin{equation}
    \pdf_{u_r}(r)  = \int \pdf_{u_r \mid \abs{\eta}}(r\mid n) \pdf_{\abs{\eta}}(n)  \,  dn \label{eq:conditionalrare},
\end{equation}
where we have already derived the pdf for $\pdf_{\abs{\eta}}$ in~\cref{eq:pdf_eta_fold}. What   remains is the derivation of the conditional pdf for $\pdf_{u_r \mid \abs{\eta}} $.

By conditioning on $\abs{\eta} = n$, we find  the derived    distribution for the  conditional pdf given by
\begin{equation}\label{eq:rareconditionalpdf}
    \pdf_{u_r \mid \abs{\eta}}(r\mid n)  = \frac{1}{r \zeta \omega_n  \tau_{e}} \biggl\{  s\left(r-\frac{n}{\omega_d} \e^{-\zeta\omega_n\tau_{e} }\right) - s\left(r-\frac{n}{\omega_d} \right)  \biggr\},
\end{equation}
where $s(\blank)$ denotes the step function, which is equal to $1$ when the argument is greater or equal to $0$ and $0$ otherwise. We refer to~\cref{sec:pdf_derivation} for a  detailed derivation.

Using the \cref{eq:rareconditionalpdf,eq:pdf_eta}, in~\cref{eq:conditionalrare} we obtain the final result for the rare event distribution for response displacement as
\begin{align}
    \pdf_{u_{r}}(r) & = \int \pdf_{u_{r}\mid \abs{\eta}}(r\mid n)  \pdf_{\abs{\eta}}(n)\, dn, \\
                    & \!\begin{multlined}                                         
    =  \frac{1}{r \zeta \omega_n  \sigma_{\abs{\eta}}\tau_{e}\sqrt{2\pi}} \int^{\infty}_0  \biggl\{\exp\biggl(-\frac{(n-\mu_\alpha )^2}{2\sigma_{\abs{\eta}}^2}\biggr)+\exp\biggl(-\frac{(n+\mu_\alpha )^2}{2\sigma_{\abs{\eta}}^2}\biggr)  \biggr\}    \\
    \times \biggl\{  s\left(r-\frac{n}{\omega_d} \e^{-\zeta\omega_n\tau_{e} }\right) - s\left(r-\frac{n}{\omega_d} \right)  \biggr\}  dn.
    \end{multlined}
    \label{eq:pdf_rare}
\end{align}

\subsubsection{Summary of results for the underdamped case}\label{sec:pdf_under}

\paragraph{Displacement Envelope}
Finally, combining the results of~\cref{sec:backgroundpdf,sec:rarepdf,sec:rareeventtransitionprob} using the total probability law,
\begin{equation}
    \pdf_u(r) = \pdf_{u_b}(r) (1-\prob_{r}) + \pdf_{u_r}(r)\prob_{r},
\end{equation}
we obtain the desired envelope distribution for the displacement of the response
\begin{multline}\label{eq:ana_sdof_dis_pdf}
    \pdf_u(r) = \frac{r}{\sigma_{x_b}^2  }\exp\biggl(-\frac{r^2}{2\sigma_{x_b}^2}\biggr)(1 - \nu_\alpha \tau_{e}) \\
    +  \frac{\nu_\alpha \tau_{e}}{r \zeta \omega_n  \sigma_{\abs{\eta}}\tau_{e}\sqrt{2\pi}} \int^{\infty}_0  \biggl\{\exp\biggl(-\frac{(n-\mu_\alpha )^2}{2\sigma_{\abs{\eta}}^2}\biggr)+\exp\biggl(-\frac{(n+\mu_\alpha )^2}{2\sigma_{\abs{\eta}}^2}\biggr)  \biggr\}    \\
    \times \biggl\{  s\left(r-\frac{n}{\omega_d} \e^{-\zeta\omega_n\tau_{e} }\right) - s\left(r-\frac{n}{\omega_d} \right)  \biggr\}  dn,
\end{multline}
where   $\tau_{e} =- \frac{1}{\zeta\omega_n}\log \rho_c$ and $s(\blank)$ denotes the step function.

\paragraph{Velocity Envelope}
Similarly, we    obtain the envelope distribution of the velocity of the system. The background dynamics distribution for velocity was obtained in~\cref{eq:env_background}.  Noting that from (\ref{eq:rare_enveloperesp}), $\dot u_r = \omega_d u_r$, the rare event pdf is modified by a constant factor
\begin{equation}
    \pdf_{\dot u}(r) = \pdf_{\dot{u}_b}(r) (1-\prob_{r}) + \omega_d^{-1}\pdf_{u_r}(r/\omega_d)\prob_{r}.
\end{equation}
The final formula  for the velocity envelope pdf is given by
\begin{multline}\label{eq:ana_sdof_vel_pdf}
    \pdf_{\dot{u}}(r) = \frac{r}{\sigma_{\dot{x}_b}^2  }\exp\biggl(-\frac{r^2}{2\sigma_{\dot{x}_b}^2}\biggr)(1 - \nu_\alpha \tau_{e}) \\
    +  \frac{\nu_\alpha }{r \zeta \omega_n  \sigma_{\abs{\eta}}\sqrt{2\pi}} \int^{\infty}_0  \biggl\{\exp\biggl(-\frac{(n-\mu_\alpha )^2}{2\sigma_{\abs{\eta}}^2}\biggr)+\exp\biggl(-\frac{(n+\mu_\alpha )^2}{2\sigma_{\abs{\eta}}^2}\biggr)  \biggr\}    \\
    \times \biggl\{  s\left(r-n \e^{-\zeta\omega_n\tau_{e} }\right) - s\left(r-n \right)  \biggr\}  dn.
\end{multline}

\paragraph{Acceleration Envelope}
Lastly we also obtain the envelope distribution of the acceleration.  Noting that from eq. (\ref{eq:rare_enveloperesp}), $\ddot u_r = \omega_d^2 u_r$, the rare event pdf for acceleration is also modified by a constant factor
\begin{equation}
    \pdf_{\ddot u}(r) = \pdf_{\ddot{u}_b}(r) (1-\prob_r) + \omega_d^{-2}\pdf_{u_r}(r/\omega_d^2)\prob_{r}.
\end{equation}
The final formula  for the acceleration envelope pdf is then
\begin{multline}\label{eq:ana_sdof_acc_pdf}
    \pdf_{\ddot{u}}(r) = \frac{r}{\sigma_{\ddot{x}_b}^2  }\exp\biggl(-\frac{r^2}{2\sigma_{\ddot{x}_b}^2}\biggr)(1 - \nu_\alpha \tau_{e}) \\
    +  \frac{\nu_\alpha }{r \zeta \omega_n  \sigma_{\abs{\eta}}\sqrt{2\pi} } \int^{\infty}_0  \biggl\{\exp\biggl(-\frac{(n-\mu_\alpha)^2}{2\sigma_{\abs{\eta}}^2}\biggr)+\exp\biggl(-\frac{(n+\mu_\alpha )^2}{2\sigma_{\abs{\eta}}^2}\biggr)  \biggr\}    \\
    \times \biggl\{  s\left(r-n\omega_d\e^{-\zeta\omega_n\tau_{e} }\right) - s\left(r-n\omega_d \right)  \biggr\}  dn.
\end{multline}

\subsubsection{Comparison with Monte-Carlo simulations}\label{sec:sodf_ana_mc_comp}

For the Monte-Carlo simulations the excitation time series is generated by superimposing the background and rare event components. The background excitation, described by a stationary stochastic process with a Pierson-Moskowitz spectrum   (\cref{eq:pmspectrum}), is simulated through  a superposition of cosines over a range of frequencies with corresponding amplitudes  and uniformly distributed random phases. The intermittent  component  is the random impulse train, and each impact is  introduced as a  velocity jump with a given magnitude at the point of the impulse impact. For each of the comparisons performed in this work we generate $10$ realizations of the excitation time series, each with a train of $100$ impulses. Once each ensemble time series for the excitation is  computed, the governing ordinary differential equations  are solved using a 4th/5th order Runge-Kutta method (we carefully  account for the modifications in the momentum that an impulse imparts by integrating up to each impulse time and modifying the initial conditions that the impulse imparts before   integrating the system to  the next impulse time). For each realization the system is integrated for a sufficiently long time so that we have converged   response statistics for the displacement, velocity, and acceleration.

\begin{figure}[H]
\centerline{
\includegraphics[width=1.25\textwidth]{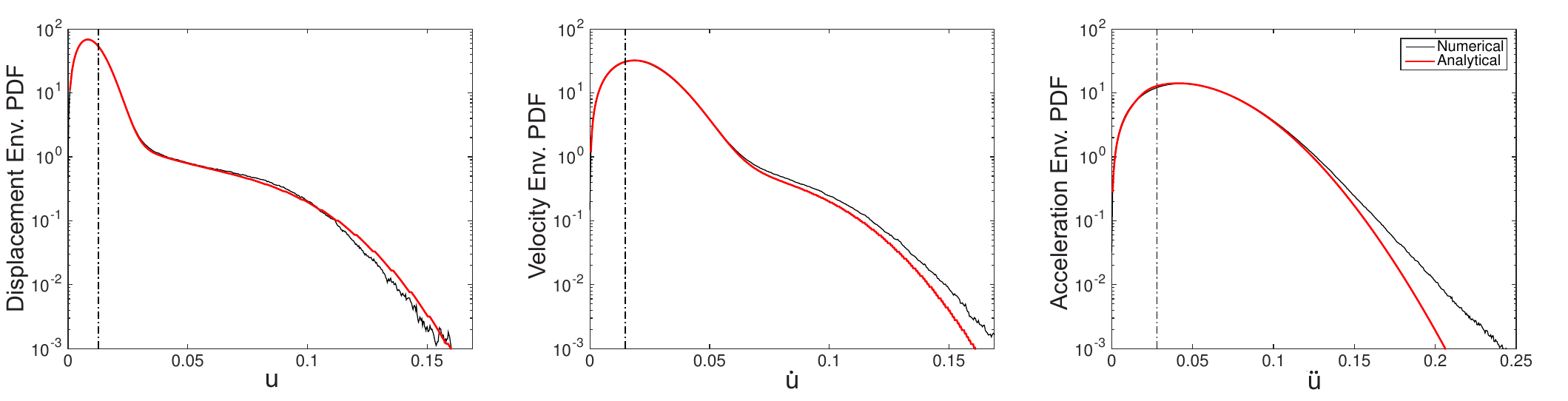}}
    \caption{\textbf{[Severely underdamped case]} Comparison between direct
Monte-Carlo simulation and the analytical pdf for the SDOF system 1. The pdf for the envelope of
each of the stochastic variables, displacement, velocity, and acceleration, are presented. The
dashed
line indicates one standard deviation. }
\label{fig:sdof1}
\end{figure}

We utilize a  shifted Pierson-Moskowitz spectrum $S_{hh}(\omega-1)$ in order to avoid resonance.
The other parameters and resulted statistical quantities of the system are given in
\cref{tab:sdof1}. As it can be seen in \cref{fig:sdof1} the analytical approximations compare
well with the Monte-Carlo simulations many standard deviations away from the zero mean. The
results are robust to different parameters as far as we satisfy the assumption of independent (non-
overlapping) random events.

Some discrepancies shown between Monte-Carlo simulations can be attributed to the envelope approximation used for the rare event quantification. Indeed, these discrepancies are reduced significantly if one utilizes the semi-analytical method presented in the next section, where we do not make any simplifications for the form of the response during extreme events.
\begin{table}[H]
\centering
\caption{Parameters and relevant statistical quantities for SDOF system 1. }
\begin{tabular}{lr|lr}
\toprule
$\lambda$     &    $ 0.01$      &      $k$      &    $1$  \\ 
$T_\alpha$    &    $5000$      &  $\zeta$    & $0.005$ \\ 
$\omega_n$  &    $1$            &  $\omega_d$  &    $1$ \\ 
$\mu_\alpha =7\times\sigma_{\dot{h}}$  &    $0.1$            &  $q$  &    $1.582\times 10^{-4}$ \\ 
$\sigma_\alpha=\sigma_{\dot{h}}$  &    $0.0143$            &  $\sigma_{h}$  &    $0.0063$ \\ 
$\sigma_{\dot{x}_b}$  &    $0.0179$     &  $\sigma_{x_b}$  &    $0.0082$ \\ 
$\sigma_{|\eta|}$  &    $0.0229$                   &  $\prob_r$  &    $0.0647$ \\ \bottomrule
\end{tabular}
\label{tab:sdof1}
\end{table}

\subsection{Analytical pdf for the overdamped case~\texorpdfstring{$\bm{\zeta\gg1}$}{zeta >> 1}}\label{sec:pdf_over}
In the  previous section, we  illustrated the derivation of the   analytical response pdf under the assumption $ \zeta\ll1 $. Here, we briefly summarize the   results for the  response pdf for the case where $\zeta\gg1$. One can follow the same steps using the corresponding formulas for the rare event transitions in the presence of large damping (\cref{sec:raretransitions}). \textit{An important difference for the overdamped case is that the system does not exhibit highly oscillatory motion as opposed to the underdamped case, and hence we directly work on the response pdf instead of the envelope pdf}.

\paragraph{Displacement} The total probability law becomes
\begin{equation}
    \pdf_x(r) = \pdf_{x_b}(r) (1-\prob_{r, \text{dis}}) + \pdf_{x_r}(r)\prob_{r, \text{dis}},
\end{equation}
and we obtain the following pdf for the displacement of the system 
\begin{multline}\label{eq:semi_sdof_dis_pdf}
    \pdf_x(r) = \frac{1}{\sigma_{x_b}\sqrt{2\pi}  }\exp\biggl(-\frac{r^2}{2\sigma_{x_b}^2}\biggr)(1 - \nu_\alpha \tau_{e, \text{dis}}) \\
+  \frac{\nu_\alpha \tau_{e, \text{dis}}}{r (\zeta \omega_n-\omega_o) \sigma_{\eta}\sqrt{2\pi}\left(\tau_{e, \text{dis}}-\tau_s\right)} \int^{\infty}_{-\infty}  \exp\biggl(-\frac{(n-\mu_\alpha )^2}{2\sigma_{\eta}^2}\biggr)\\
    \times\biggl\{  s\left(r-\frac{n}{2\omega_o} \e^{-(\zeta\omega_n-\omega_o)\tau_{e, \text{dis}} }\right) - s\left(r-\frac{n}{2\omega_o} \right)  \biggr\} dn,
\end{multline}
where  $\tau_{e, \text{dis}} =  \frac{\pi}{2\omega_o} - \frac{1}{\zeta \omega_n-\omega_o}\log \rho_c$.

\paragraph{Velocity}
Similarly we derive the total probability law for the response velocity
\begin{equation}
    \pdf_{\dot x}(r) = \pdf_{\dot{x}_b}(r) (1-\prob_{r, \text{vel}}) + \pdf_{\dot{x}_r}(r)\prob_{r, \text{vel}}.
\end{equation}
The final result   for the velocity  pdf is 
\begin{multline}\label{eq:semi_sdof_vel_pdf}
    \pdf_{\dot{x}}(r) = \frac{1}{\sigma_{\dot{x}_b}\sqrt{2\pi}  }\exp\biggl(-\frac{r^2}{2\sigma_{\dot{x}_b}^2}\biggr)(1 - \nu_\alpha \tau_{e, \text{vel}}) \\
   +  \frac{\nu_\alpha \tau_{e, \text{vel}}}{r (\zeta \omega_n+\omega_o)   \sigma_{\eta}\sqrt{2\pi} \tau_{e, \text{vel}}} \int^{\infty}_{-\infty}  \exp\biggl(-\frac{(n-\mu_\alpha )^2}{2\sigma_{\eta}^2}\biggr) \\
    \times  \biggl\{  s\left(r-n \e^{-(\zeta \omega_n+\omega_o)\tau_{e, \text{vel}} }\right) - s\left(r-n \right)  \biggr\} dn,
\end{multline}
where $\tau_{e, \text{vel}} = - \frac{1}{\zeta \omega_n+\omega_o}\log \rho_c$.

\paragraph{Acceleration}
The total probability law for the response acceleration is
\begin{equation}
    \pdf_{\ddot x}(r) = \pdf_{\ddot{x}_b}(r) (1-\prob_{r, \text{acc}}) + \pdf_{\ddot{x}_r}(r)\prob_{r, \text{acc}},
\end{equation}
and this gives the following result for the acceleration pdf
\begin{multline}\label{eq:semi_sdof_acc_pdf}
    \pdf_{\ddot{x}}(r) = \frac{1}{\sigma_{\ddot{x}_b}\sqrt{2\pi}  }\exp\biggl(-\frac{r^2}{2\sigma_{\ddot{x}_b}^2}\biggr)(1 - \nu_\alpha \tau_{e, \text{acc}}) \\
   +  \frac{\nu_\alpha \tau_{e, \text{acc}}}{r (\zeta \omega_n+\omega_o)   \sigma_{\eta}\sqrt{2\pi}  \tau_{e, \text{acc}}} \int^{\infty}_{-\infty}  \exp\biggl(-\frac{(n-\mu_\alpha )^2}{2\sigma_{\eta}^2}\biggr)  \\
    \times  \biggl\{  s\left(r-n(\zeta \omega_n+\omega_o)\e^{-(\zeta \omega_n+\omega_o)\tau_{e, \text{acc}} }\right) - s\left(r-n\left(\zeta \omega_n+\omega_o\right)\right)  \biggr\} dn,
\end{multline}
where $\tau_{e, \text{acc}} = - \frac{1}{\zeta \omega_n+\omega_o}\log \rho_c$. Note that in this case we do not have the simple scaling as in  the underdamped case for the conditionally rare pdf.

\subsubsection{Comparison with Monte-Carlo simulations}\label{sec:sodf_semi_mc_comp}

\begin{figure}[H]
    \centering
    \centerline{\includegraphics[width=1.25\textwidth]{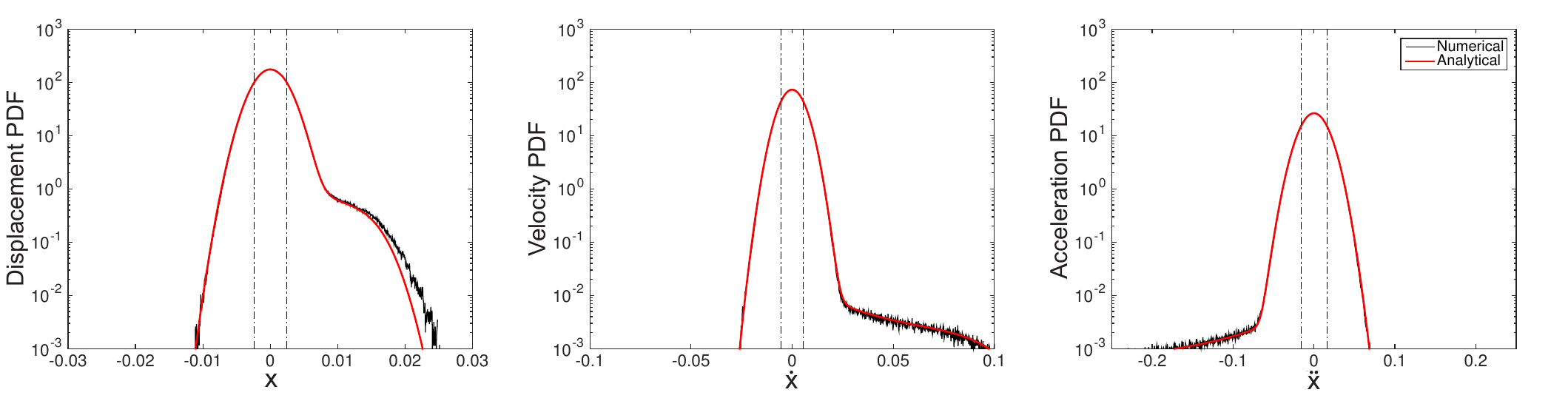}}
    \caption{\textbf{[Severely overdamped case]} Comparison between direct
Monte-Carlo simulation and the analytical pdf for SDOF system 2. The pdf for the value of each stochastic process is shown.  The dashed line indicates one standard deviation. }
\label{fig:sdof2}
\end{figure}

We confirm the accuracy of the  analytical results  given in~\cref{eq:semi_sdof_dis_pdf,eq:semi_sdof_vel_pdf,eq:semi_sdof_acc_pdf} for the strongly overdamped case through comparison with direct Monte-Carlo simulations. The parameters and resulted statistical quantities of the system are given in \cref{tab:sdof2}. The analytical estimates show favorable agreement  with numerical simulations for this case (\cref{fig:sdof2}), just as in the previous underdamped case.

\begin{table}[H]
\centering
\caption{Parameters and relevant statistical quantities for SDOF system 2. }
\begin{tabular}{lr|lr}
\toprule
$\lambda$     &    $ 6$      &      $k$      &    $1$  \\ 
$T_\alpha$    &    $1000$      &  $\zeta$    & $3$ \\ 
$\omega_n$  &    $1$            &  $\omega_d$  &    $2.828$ \\ 
$\mu_\alpha =7\times\sigma_{\dot{h}}$  &    $0.1$            &  $q$  &    $1.582\times 10^{-4}$ \\ 
$\sigma_\alpha=\sigma_{\dot{h}}$  &    $0.0143$            &  $\sigma_{h}$  &    $0.0063$ \\ 
$\sigma_{\dot{x}_b}$  &    $0.0056$            &  $\sigma_{x_b}$  &    $0.0022$ \\ 
$\sigma_{\eta}$  &    $0.0154$            &  $\prob_{r, dis}$  &    $0.0140$ \\ 
 $\prob_{r, vel}$  &    $0.0004$            &  $\prob_{r, acc}$  &    $0.0004$ \\ 
 \bottomrule
\end{tabular}
\label{tab:sdof2}
\end{table}

\section{Semi-analytical pdf of the response of SDOF systems}\label{sec:sdof_semi}
We now formulate a semi-analytical approach to quantify the response pdf for any arbitrary set of
system parameters, including the severely underdamped  or overdamped cases  considered previously.
The approach here adapts the numerical scheme described in~\cite{mohamad2016b} for  systems
undergoing internal instabilities.

While for the limiting cases that we studied previously knowledge of the trajectory (time series)
of the system ($x_{r}(t)$ or $u_{r}(t)$) could analytically be translated to information about the
corresponding pdf ($f_{x_r}$ or $f_{u_r}$), this is not always possible. In addition, for
\textit{nonlinear} structural systems one will not, in general, have analytical expressions for the
rare event transitions. For these cases we can compute the rare event statistics by numerically
approximating the corresponding histogram, using either analytical or numerically generated
trajectories for the rare event regime.

\subsection{Numerical computation of rare events statistics}\label{sec:num_hist}

Consider the same SDOF system introduced in~\cref{sec:sdof_ana}.  Recall that we have quantified the response pdf by the PDS method using  the total probability law
\begin{equation}
    \pdf_x(r) = \pdf_{x_b}(r) (1-\prob_{r}) + \pdf_{x_r}(r)\prob_{r}.
\end{equation}
In the  previous section, the  derivation consisted of estimating all three unknown quantities: the background distribution $\pdf_{x_b}$, the rare event distribution $\pdf_{x_r}$, and the rare event probability $\prob_{r}$  analytically. However, in the semi-analytical scheme we will obtain  the rare event distribution $ \pdf_{x_r}$ and rare event probability $\prob_{r}$  by taking a histogram of the numerically simulated analytical form of the rare response. The background distribution $\pdf_{x_b}$ is still  obtained analytically as in~\cref{sec:backgroundpdf}.

Recall that the rare event distribution is given by
\begin{align}
    \pdf_{x_{r}}(r) & = \int \pdf_{x_{r}\mid \eta}(r\mid n)  \pdf_{\eta}(n)\, dn, 
\end{align}
where $\pdf_{\eta}(n)$ is known  analytically (\cref{eq:pdf_eta}). It is     the conditional pdf $\pdf_{x_{r}\mid \eta}(r\mid n) $ that   we estimate by a  histogram:
\begin{equation}
    f_{x_r \mid \eta}(r\mid n)  = \hist\bigl\{ x_{r\mid \eta}(t\mid n) \bigr\},  \quad   t = [0,  \tau_{e, \text{dis}}],
\end{equation}
where we use the analytical solution of the oscillator with non-zero initial velocity, $n$:
\begin{align}
    x_{r\mid \eta}(t\mid n) =  \frac{n}{2\omega_o}  \biggl(\e^{-(\zeta\omega_n-\omega_o) t}- \e^{-(\zeta\omega_n+\omega_o)  t}\biggr). 
\end{align}
The histogram is taken from $t=0$ (the beginning of the rare event) until the end of the rare  event at $t=\tau_e$.   The conditional distribution of rare event response for the velocity and acceleration can   be written as well:
\begin{align}
    f_{\dot{x}_r \mid \eta}(r\mid n)  & = \hist\bigl\{ \dot{x}_{r \mid \eta}(t \mid n) \bigr\}, \quad   t = [0,   \tau_{e, \text{vel}}], \\
    f_{\ddot{x}_r \mid \eta}(r\mid n) & = \hist\bigl\{ \ddot{x}_{r \mid \eta}(t \mid n) \bigr\}, \quad  t = [0,   \tau_{e, \text{acc}}].
\end{align}

\subsection{Numerical estimation of the rare events probability} \label{sec:num_dur}

In order to compute the histogram of a  rare impulse event, the duration of a rare response needs to be obtained  numerically. Recall that we have defined the duration of  a rare responses  by
\begin{equation}
    x_r(\tau_{e})=  \rho_c \, \max\big\{|x_r|\big\},
\end{equation}
where $\rho_c = 0.1$. In the numerical computation of $\tau_e$ the absolute maximum of the response needs to be estimated numerically as well. Once the rare event duration has been specified, we can obtain the probability of  rare events by
\begin{equation}
    \prob_{r}= \nu_\alpha \tau_{e} = \tau_{e}/ T_\alpha.
\end{equation}
This value is     independent of the conditional background magnitude. The above procedure is applied for the rare event response displacement $\tau_{e, \text{dis}}$, velocity $\tau_{e, \text{vel}}$, and acceleration $\tau_{e, \text{acc}}$.

\subsection{Semi-analytical probability density functions}

We can now compute the  response pdf using the described semi-analytical approach. For the
displacement we have:
\begin{align}
    \pdf_x(r) = \frac{1 - \nu_\alpha \tau_{e, \text{dis}}}{\sigma_{x_b}\sqrt{2\pi}  }\exp\biggl(-\frac{r^2}{2\sigma_{x_b}^2}\biggr) +  \nu_\alpha \tau_{e, \text{dis}} \int^\infty_0 \hist\bigl \{ x_{r\mid \eta}(t\mid n) \bigr \} \pdf_{\eta}(n)\, dn. 
\end{align}
The corresponding pdf for the velocity $\pdf_{\dot{x}}$ and acceleration $\pdf_{\ddot{x}}$, can be computed with the same formula but with the appropriate variance for the Gaussian core ($\sigma_{\dot{x}_b}$ or $\sigma_{\ddot{x}_b}$), rare event duration ($\tau_{e,
\text{vel}}$ or $\tau_{e,
\text{acc}}$), and histograms ($\dot{x}_{r\mid \eta}(t\mid n)$ or $\ddot{x}_{r\mid \eta}(t\mid n)$).

\subsubsection{Comparison with Monte-Carlo simulations}

\begin{figure}[H]
    \centering
    \centerline{\includegraphics[width=1.25\textwidth]{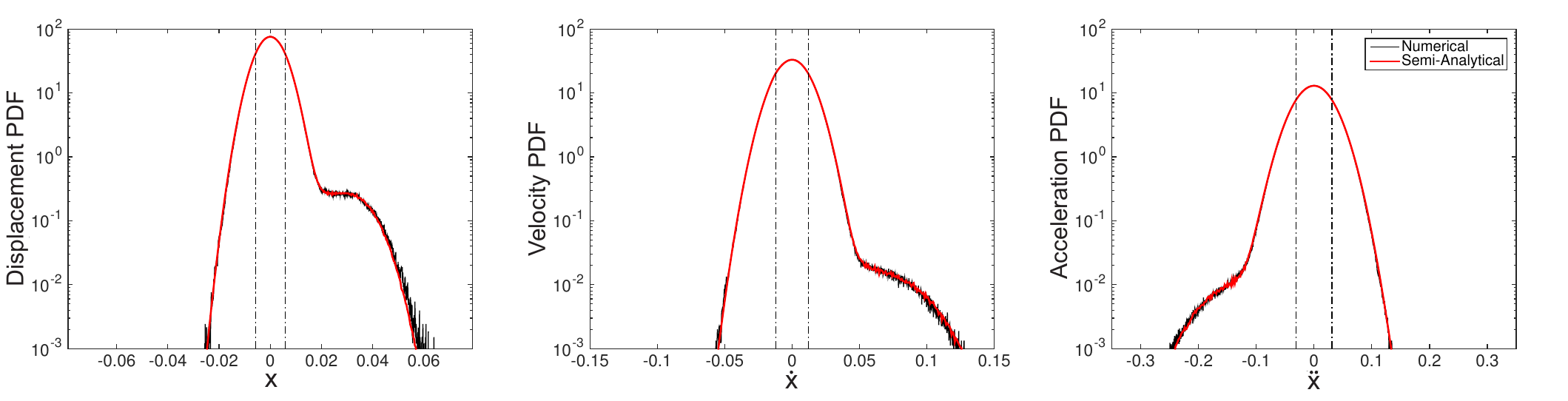}}
    \caption{\textbf{[Critically damped system]} Comparison between direct Monte-Carlo simulations and the semi-analytical pdf for SDOF system 3.  Dashed lines indicate one standard deviation. }
    \label{fig:sdof3}
\end{figure}

For illustration, a SDOF configuration is considered with  critical damping ratio,  $\zeta=1$. The
detailed parameters and relevant statistical quantities of the system are given in~\cref{tab:sdof3}.
This is a regime where  the  analytical results derived in~\cref{sec:sdof_ana} are not applicable.
Even for this $\zeta$ value, the semi-analytical  pdf for the response shows excellent agreement 
with   direct simulations  (\cref{fig:sdof3}). We emphasize that the computational cost of the
semi-analytical scheme is comparable with that of the analytical approximations (order of
seconds) and both are significantly lower than the cost of Monte-Carlo simulation (order of hours).

\begin{table}[H]
\centering
\caption{Parameters and relevant statistical quantities for SDOF system 3. }
\begin{tabular}{lr|lr}
\toprule
$\lambda$     &    $ 2$      &      $k$      &    $1$  \\ 
$T_\alpha$    &    $400$      &  $\zeta$    & $1$ \\ 
$\omega_n$  &    $1$            &  $\omega_d$  &    $0$ \\ 
$\mu_\alpha =7\times\sigma_{\dot{h}}$  &    $0.1$            &  $q$  &    $1.582\times 10^{-4}$ \\ 
$\sigma_\alpha=\sigma_{\dot{h}}$  &    $0.0143$            &  $\sigma_{h}$  &    $0.0063$ \\ 
$\sigma_{\dot{x}_b}$  &    $0.0120$            &  $\sigma_{x_b}$  &    $0.0052$ \\ 
$\sigma_\alpha=\sigma_{\eta}$  &    $0.0187$            &  $\prob_{r, dis}$  &    $0.0122$ \\ 
 $\prob_{r, vel}$  &    $0.0075$            &  $\prob_{r, acc}$  &    $0.0032$ \\ 
 \bottomrule
\end{tabular}
\label{tab:sdof3}
\end{table}

\section{Semi-analytical pdf for the response of MDOF systems}\label{sec:2dof_semi}

\begin{figure}[H]
    \centering
    \includegraphics[height=0.225\textwidth]{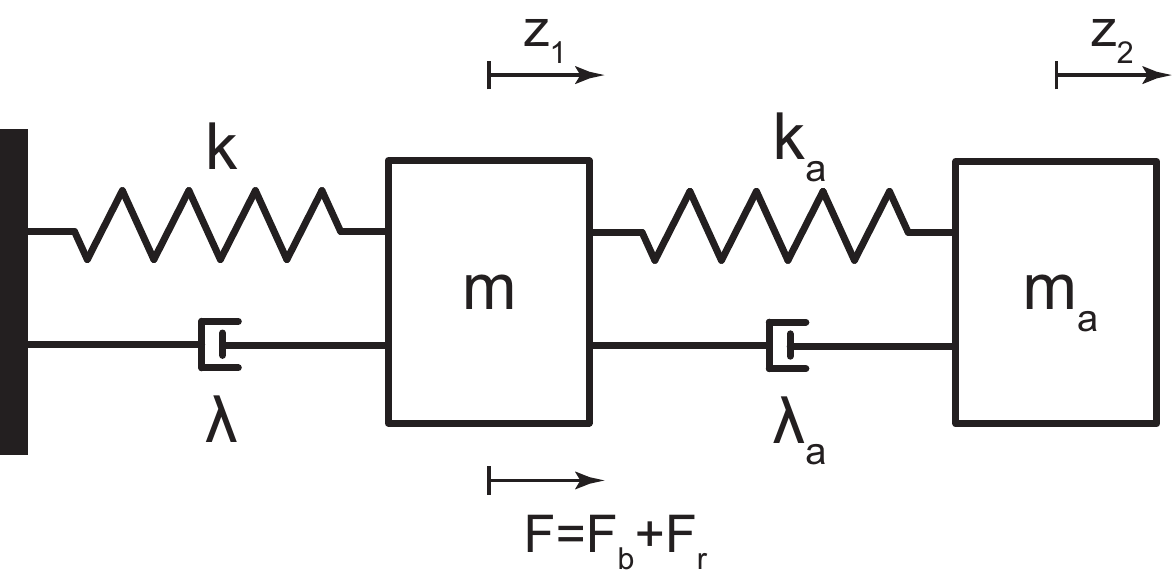}
    \caption{The considered  TDOF system. The excitation is applied to the first mass,
    with mass $m$.}
    \label{fig:prot2}
\end{figure}

An important advantage of the semi-analytical scheme is the straightforward applicability of the
algorithm to MDOF   systems. In this section we demonstrate how the extension can be made for a 
two-degree-of-freedom (TDOF)  linear system (see~\cref{fig:prot2}): 
\begin{align}\label{eq:2dofsys}
    m\ddot{x} +  \lambda \dot{x} + k x +  \lambda_a (\dot{x}-\dot{y})  + k_a(x-y) & = F(t), \\
    m_a\ddot{y}+ \lambda_a (\dot{y}-\dot{x}) +k_a (y-x)                           & =   0,  
\end{align}
where the stochastic forcing $F(t) = F_b(t)+F_r(t)$ is applied to the first mass (mass $m$)
and $x,y$ are   displacements of the two masses. As before,  $F_b
(t) $ is the background component and $F_r(t) = 
\sum^{N(t)}_{i=1} \alpha_i \,
\delta(t-\tau_i)$ is the rare event component.

The background statistics are obtained by analyzing the response spectrum of the TDOF system subjected to the background excitation component. The details are given in Appendix \ref{sec:backstat}. For a nonlinear system this can be done using statistical linearization method \cite{Rob_SPanos03}. The histograms for the rare event transitions can be computed through standard analytical expressions that one can derive for a linear system like the one we consider under the set of initial conditions: $x_0=y_0=\dot y_0 =0$ and $\dot x_0=n.$

Once the impulse response has been obtained we numerically quantify the rare event distribution, as well as the rare event duration and use the semi-analytical decomposition. The pdf  is then given by
\begin{align}              
    \pdf_z(r) = \frac{1 - \nu_\alpha \tau_{e}^z}{\sigma_{z_b}\sqrt{2\pi}  }\exp\biggl(-\frac{r^2}{2\sigma_{z_b}^2}\biggr) +  \nu_\alpha \tau_{e}^z \int^\infty_0 \hist\bigl\{ z_{r\mid \eta}(t\mid n) \bigr\} \pdf_{\eta}(n)\, dn,
\end{align}
where $z$ can be  either of the degrees-of-freedom $(x$ or $y)$ or the corresponding velocities or accelerations, while $\tau_{e}^z$ is the typical duration of the rare events and is estimated numerically. We note that as in the previous cases the pdf is composed of a Gaussian core (describing the background statistics) as well as, a heavy tailed component that is connected with the rare transitions.  For each case of $z$ the corresponding variance under background excitation, temporal durations of rare events, and histograms for rare events should be employed.

\begin{figure}[H]
    \centering
    \centerline{\includegraphics[width=1.25\textwidth]{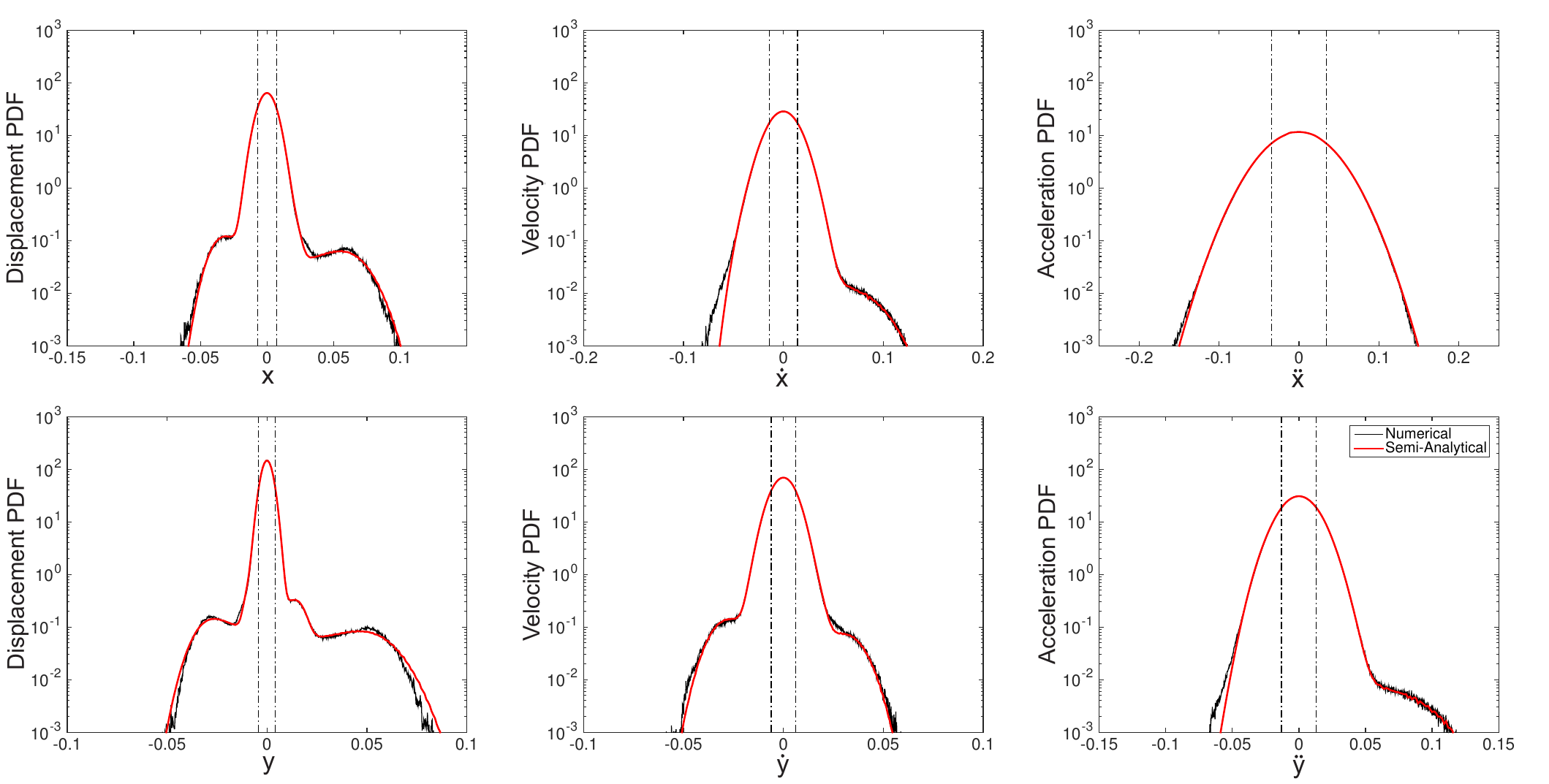}}
    \caption{\textbf{[Two DOF System]} Comparison between direct Monte-Carlo simulation and the semi-analytical approximation. The pdf for the value of the time series are presented.  Dashed line indicates one standard deviation. }
    \label{fig:tdof}
\end{figure}

Results are presented for the pdf of the  displacement, velocity and acceleration of each degree-of-freedom (\cref{fig:tdof}). These compare favorably with the direct Monte-Carlo simulations. The parameters and resulted statistical quantities of the system are given in \cref{tab:tdof}. Further numerical simulations (not presented)  demonstrated strong robustness of the approach.

\begin{table}[H]
\centering
\caption{Parameters and relevant statistical quantities for the TDOF system. }
\begin{tabular}{lr|lr}
\toprule
$m$     &    $ 1$      &      $m_a$      &    $1$  \\ 
$\lambda$     &    $ 0.01$      &      $k$      &    $1$  \\ 
$\lambda_a$     &    $ 1$      &      $k_a$      &    $0.1$  \\ 
$T_\alpha$    &    $1000$      &   $\sigma_{\eta}$  &    $0.0199$      \\ 
$\mu_\alpha$  &    $0.1$            &  $q$  &    $1.582\times 10^{-4}$ \\ 
$\sigma_\alpha$  &     $0.0143$          &  $\sigma_{F_b}$  &   $0.0351$  \\ 
$\prob_{r, dis}^x$  &    $0.0177$   &     $\prob_{r, dis}^y$  &    $0.0190$ \\ 
$\prob_{r, vel}^x$  &    $0.0098$   &     $\prob_{r, vel}^y$  &    $0.0209$ \\ 
$\prob_{r, acc}^x$  &    $0.0066$   &     $\prob_{r, acc}^y$  &    $0.0082$ \\
\bottomrule
\end{tabular}
\label{tab:tdof}
\end{table}

\section{Semi-analytical pdf of local maxima}

It is straightforward to extend the semi-analytical framework for other quantities of interest,
such as the local extrema (maxima and minima) of the response. The numerically generated histogram
for the rare transitions can be directly computed for the local extrema as well. On the other hand,
for the background excitation regime, we use known   results from the theory of stationary Gaussian
stochastic processes to describe  the corresponding background pdf analytically.

\subsection{Distribution of local maxima under background excitation}

For a stationary Gaussian process with arbitrary spectral bandwidth $\epsilon$, the probability
density function of positive extrema (maxima) is given by~\cite{Ochi90, Karadeniz12,}:
\begin{equation}
f_{m^+} (\xi) = \frac{\epsilon}{\sqrt{2\pi}}e^{-\xi^2/2\epsilon^2}+\sqrt{1-\epsilon^2}\xi e^
{-\xi^2/2}\Phi\biggl(\frac{\sqrt{1-\epsilon^2}}{\epsilon}\xi\biggr),  \quad -\infty \leq \xi
\leq \infty,
\label{eqn:pdf_max}
\end{equation}
where $\xi = \frac{x}{\sqrt{\mu_0}}$, $x$ is the magnitude of the maxima, spectral bandwidth $\epsilon = \sqrt{1-\frac{\mu_2^2}{\mu_0\mu_4}}$, and $\Phi(x) = \frac{1}{\sqrt{2\pi}}\int^x_{-\infty} e^{-u^2/2}du$ is the standard normal cumulative distribution function.
The spectral moments for the background response displacement $x_b$ are also defined as
\begin{equation}
\mu_n = \int^\infty_{0} \omega^n S_{x_bx_b}(\omega)\,d\omega.
\end{equation}
We note that that for the limit of an infinitesimal narrow-banded signal $(\epsilon = 0)$, the pdf
converges to a Rayleigh distribution. On the other hand for an infinitely broad-banded signal $(\epsilon=1)$, the distribution converges to the Gaussian pdf. For a signal with in-between spectral bandwidth $(0\leq \epsilon\leq 1)$, the pdf has a blended structure with the form in~\cref{eqn:pdf_max}. 

Taking into account the asymmetric nature of the intermittent excitation, we need to consider both  the positive and negative extrema. For  the background excitation the pdf can be written as:
\begin{equation}
f_{\hat{x}_b} (x) =\frac{1}{2\sqrt{\mu_0}} \Bigl\{ f_{m^+}\Bigl(\frac{x}{\mu_0}\Bigr) +  f_{m^+}
\Bigl(-\frac{x}{\mu_0}\Bigr) \Bigr\}, \quad -\infty \leq x \leq \infty,
\end{equation}
where the $\hat x$ notation denotes the local extrema of $x$. Similar expressions can be
obtained
for the velocity and acceleration extrema.

\subsection{Statistics of local extrema during rare transitions}

The conditional pdf $\pdf_{\hat{x}_{r}\mid \eta}$ for the local extrema can be numerically estimated through the histogram:
\begin{equation}
    f_{\hat{x}_r \mid \eta}(r\mid n)  = \hist\bigl\{\mathcal{M}\left( x_{r\mid \eta}(t\mid n) \right) \bigr\},  \quad   t = [0,  \tau_{e, \text{dis}}],
\end{equation}
where $\mathcal{M}(\blank)$ is an operator that gives all the positive/negative extrema. The
positive/negative extrema are defined as the points where the  derivative of the signal is
zero.

\subsection{Semi-analytical probability density function for local extrema}

\begin{figure}[H]
    \centering
    \centerline{\includegraphics[width=1.25\textwidth]{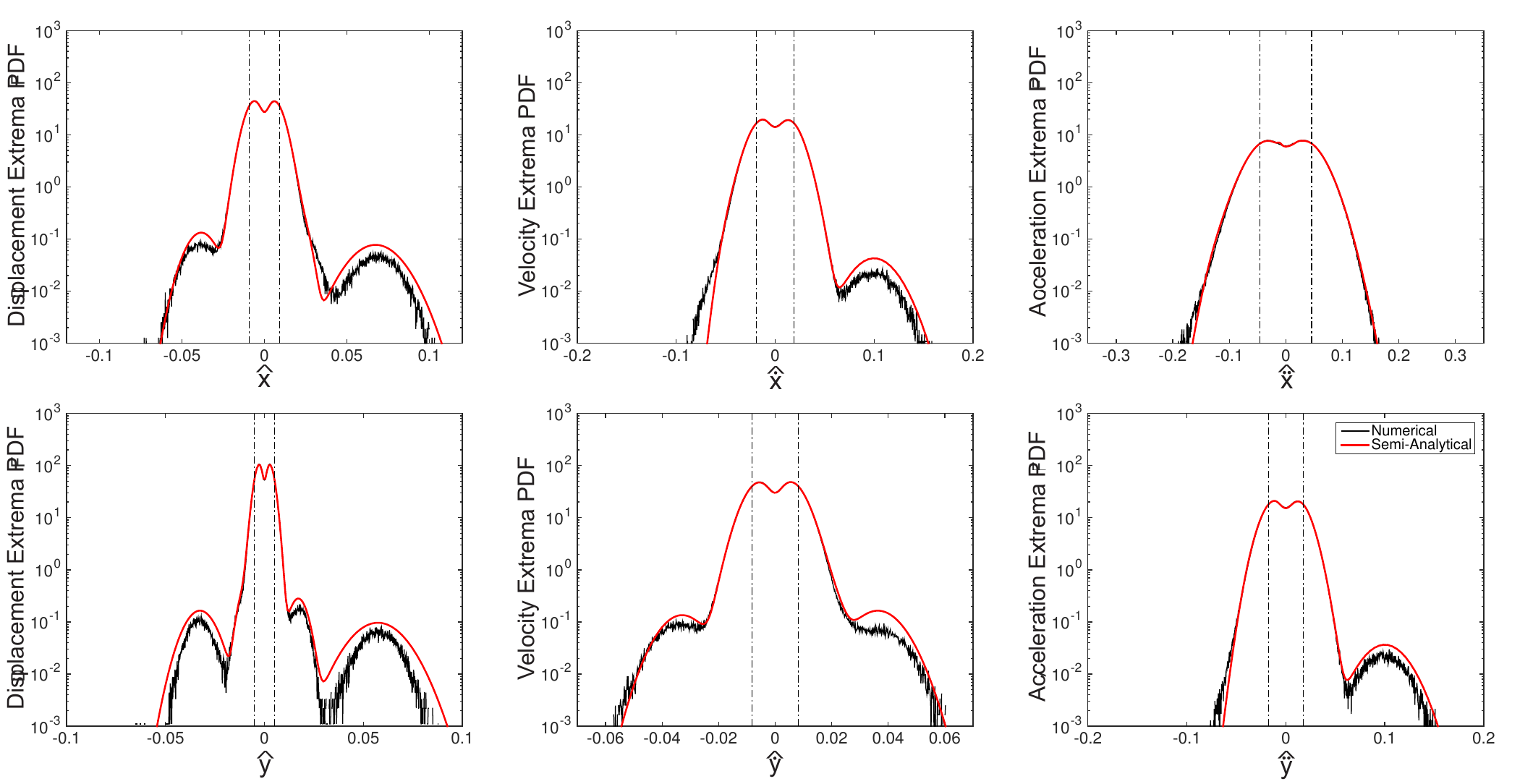}}
    \caption{\textbf{[Local extrema for TDOF]} Comparison between direct Monte-Carlo
simulation and the semi-analytical approximation. The pdf for the local extrema
of the response are presented.  Dashed line indicates one standard deviation. }
\label{fig:tdof_peak}
\end{figure}

The last step consists of applying the decomposition of the pdf. This takes the form: 
\begin{equation}
    f_{\hat{x}}(r) = \left(1 - \nu_\alpha \tau_{e, \text{dis}}\right)f_{\hat{x}_b}(r )+  \nu_\alpha \tau_{e, \text{dis}} \int^\infty_0 \text{Hist}\bigl \{ \mathcal{M}\left( x_{r\mid \eta}(t\mid n)\right) \bigr \} f_{\eta}(n)\, dn. 
\end{equation}
The same decomposition can be used for the velocity and acceleration local maxima.
We compare the semi-analytical decomposition with Monte-Carlo simulations. In \cref{fig:tdof_peak} we present results for the two-degree-of-freedom system. The pdf are shown for local extrema  of the displacements, velocities and accelerations for each degree of freedom. We emphasize the non-trivial structure of the pdf and especially their tails. Throughout these comparisons the semi-analytical scheme demonstrates accurate estimation of both the heavy tails and the non-Gaussian/non-Rayleigh structure of the background local extrema distribution.

\section{Summary and conclusions}\label{sec:conc}

We have formulated a   robust approximation method to quantify the probabilistic  response of structural systems subjected to stochastic excitation containing intermittent components. The foundation of our approach is the  recently developed  probabilistic decomposition-synthesis method for the quantification of rare events due to internal instabilities   to the problem where extreme responses are triggered by   external   forcing. The  intermittent forcing is represented as a  background component, modeled through a colored   processes with energy distributed across a range of frequencies, and additionally a rare/extreme component that can be   represented by impulses that are Poisson distributed with large inter-arrival time. Owing to the nature of the forcing, even the probabilistic response of  a linear system can be  highly complex with asymmetry and complicated  tail behavior  that is  far from Gaussian, which is  the expected form of the response pdf if the forcing did not contain an intermittently extreme component. 

The main result of this work is the derivation of  analytical/semi-analytical expressions for the
pdf of the response and its local extrema for structural systems (including the response
displacement, velocity, and acceleration pdf). These expressions decompose the pdf into a
probabilistic core, capturing the statistics under background excitation, as well as a heavy-tailed
component associated with the extreme transitions due to  the rare impacts. We have performed a
thorough analysis for linear SDOF systems under various system parameters and also derived
analytical formulas for  two special cases of parameters (lightly damped or heavily damped
systems). The general semi-analytical decomposition is applicable for any arbitrary set of system
parameters and we have demonstrated its validity through comprehensive  comparisons with Monte-
Carlo simulations. The general framework is also directly applicable to multi-degree-of-freedom
MDOF systems, as well as systems with nonlinearities, and we have assessed its performance through a 2DOF linear system  of two coupled oscillators excited through the first mass.
Modifications of the method to compute statistics of local extrema have also been presented.

The developed approach allows for computation of the response pdf of structural systems many orders
of magnitude faster than a direct Monte-Carlo simulation, which is currently the only reliable tool
for such computations. The rapid evaluation of response pdfs for systems excited by extreme forcing
by the method presented in this work paves the way for enabling robust design of structural systems
subjected to extreme events of a  stochastic nature. Our future endeavors include the application
of the developed framework to the optimization of engineering systems where extreme event
mitigation is required. In such cases, it is usually not feasible to run Monte-Carlo simulations
for various parameter sets during the design process owing to the computational costs associated
with  low probability rare events,  let alone perform parametric optimization.  We believe our
approach is well suited to such problems and can prove to be an important  method for engineering
design and reliability assessment.

\section*{Acknowledgments}
T.P.S. has been supported through the ONR\ grants N00014-14-1-0520 and N00014-15-1-2381 and the
AFOSR grant FA9550-16-1-0231. H.K.J. and M.A.M. have been supported through the first and third
grants as graduate students. We are also grateful to the Samsung Scholarship Program for support of
H.K.J. as well as the MIT Energy Initiative for support under the grant ‘Nonlinear Energy
Harvesting From Broad-Band Vibrational Sources By Mimicking Turbulent Energy Transfer Mechanisms'.

\appendix
\section{Impulse response of SDOF systems}\label{sec:raretransitions}

The response of the
system  
\begin{equation}
    \ddot{x}_r(t) + \lambda \dot {x}_r(t) + k x_r(t) = 0,
\end{equation}
under an impulse $\alpha$ at an arbitrary time $t_0$, say  $t_0 = 0$, and
 with zero initial value but nonzero initial velocity ($(x_r,\dot x_r) = (0,\dot x_{r0})$ at $t=0^-$)   are
given by the following equations under the two limiting cases of damping:

\paragraph{Severely underdamped case $\bm{\zeta\ll1}$}
With the approximation of $\zeta\ll1$ (or $\omega_d \approx \omega_n$), we
can simplify responses as
\begin{align} \label{eq:under_res}
x_r(t) & = \frac{\alpha+\dot x_{r0}}{\omega_d} \e^{-\zeta \omega_n t} \sin \omega_d t, \quad
\\ 
\dot{x}_r(t) & = (\alpha+\dot x_{r0}) \e^{-\zeta \omega_n t} \cos \omega_d t, \quad \\
 \ddot{x}_r(t) & = - (\alpha+\dot x_{r0}) \omega_d \e^{-\zeta \omega_n t} \sin \omega_d
t.      
\end{align}

\paragraph{Severely overdamped case $\bm{\zeta\gg1}$}
Similarly, with the approximation of $\zeta\gg1$ (or $\omega_o \approx \zeta
\omega_n$), we can simplify responses as

\begin{align} \label{eq:over_res}
    x_r(t) & =\frac{(\alpha+\dot x_{r0})}{2\omega_o} \e^{-(\zeta \omega_n-\omega_o)t},\\
    \dot{x}_r(t) & = (\alpha+\dot x_{r0}) \e^{-(\zeta\omega_n+\omega_o)t},\\
    \ddot{x}_r(t) & = -(\zeta\omega_n+\omega_o) (\alpha+\dot x_{r0}) \e^{-(\zeta\omega_n+\omega_o)t}.
\end{align}

\section{Probability distribution  for an arbitrarily exponentially decaying
function}\label{sec:pdf_derivation}

Consider an arbitrary time series in the following form:
\begin{equation}
    x(t) = \mathcal{A} \e^{-\alpha t},\quad \text{where } t \sim \text{Uniform}\left(\tau_1,
\tau_2 \right),
\end{equation}
where $\mathcal{A}, \alpha >0$ and $\tau_1<\tau_2$. The cumulative distribution
function (cdf) of $x(t)$ is  
\begin{align}
    F_x(x) & = \prob(\mathcal{A} \e^{-\alpha t} < x),                   
                \\
           & = \prob\Bigl( t>\frac{1}{\alpha}\log( \mathcal{A}/x  )\Bigr),
              \\
           & =   1- \prob\Bigl( t<\frac{1}{\alpha}\log(\mathcal{A} /{x})\Bigr),
         \\
           & =  1- \int^{\frac{1}{\alpha}\log(\mathcal{A}/x)}_{-\infty} \pdf_T(t)
\, dt, 
\end{align}
where $\pdf_T(t)$ is expressed using the  step function $s(\blank)$ as:
\begin{equation}
    \pdf_T(t) = \frac{1}{\tau_2-\tau_1} \bigl\{s(t-\tau_1)-s(t-\tau_2)\bigr\},
\quad  \tau_1 < \tau_2.
\end{equation}
The pdf of the response $x(t)$ can then  be derived by   differentiation.
\begin{align}
    \pdf_x(x) = & \frac{d}{dx}F_x(x),                                   
                                                                      \\
    =           & \frac{1}{\alpha x} \pdf_T\Bigl(\frac{1}{\alpha}\log( \mathcal{A}
{x} )\Bigr),                                                \\
    =           & \frac{1}{\alpha x(\tau_2-\tau_1)} \Bigl\{s(x-\mathcal{A}
\e^{-\alpha \tau_2}) - s(x-\mathcal{A} \e^{-\alpha \tau_1})\Bigr\}. 
\end{align}
We utilize the above formula for deriving analytical pdfs. Note that the
step function with respect to $x$ in the above has been derived using
\begin{align}
    \tau_1  < t                         & < \tau_2,                     
  \\
    -\alpha \tau_2  < -\alpha t         & <-\alpha \tau_1,              
  \\
    \mathcal{A}\e^{-\alpha \tau_2}  < x & <\mathcal{A}\e^{-\alpha \tau_1}.
\end{align}

\section{Background response for TDOF system}\label{sec:backstat}
Consider the statistical response of the system to the background forcing component,
\begin{align}
    m\ddot{x}_b +  \lambda \dot{x}_b + k x_b +  \lambda_a (\dot{x}_b-\dot{y}_b)  + k_a(x_b-y_b) = & \ F_b(t),  \\
    m_a\ddot{y}_b+ \lambda_a (\dot{y}_b-\dot{x}_b) +k_a (y_b-x_b) =                             & \  0. \nonumber 
\end{align}
The spectral densities are given by
\begin{align}
    S_{x_bx_b}(\omega)               & = \frac{\omega^4S_{F_b}(\omega)}{\bigl\{\mathcal{A}(\omega)-\frac{\mathcal{B}(\omega)^2}{\mathcal{C}(\omega)}\bigr\}\bigl\{\mathcal{A}(-\omega)-\frac{\mathcal{B}(-\omega)^2}{\mathcal{C}(-\omega)}\bigr\}} ,                                   \\ 
    S_{\dot{x}_b\dot{x}_b}(\omega)   & = \omega^2 S_{x_bx_b}(\omega),                                                                                                                                                                                                                 \\ 
    S_{\ddot{x}_b\ddot{x}_b}(\omega) & =\omega^4 S_{x_bx_b}(\omega),                                                                                                                                                                                                                  \\
    S_{y_b y_b}(\omega)               & = \frac{\omega^4S_{F_b}(\omega)}{\bigl\{\frac{\mathcal{A}(\omega)\mathcal{C}(\omega)}{\mathcal{B}(\omega)}-\mathcal{B}(\omega)\bigr\}\bigl\{\frac{\mathcal{A}(-\omega)\mathcal{C}(-\omega)}{\mathcal{B}(-\omega)}-\mathcal{B}(-\omega)\bigr\}}, \\
    S_{\dot{y}_b\dot{y}_b}(\omega)   & = \omega^2 S_{y_by_b}(\omega),                                                                                                                                                                                                                 \\
    S_{\ddot{y}_b\ddot{y}_b}(\omega) & = \omega^4 S_{y_by_b}(\omega),                                                                                                                                                                                                                 
\end{align}
where $S_{F_b}(\omega)$ is the spectral density of $F_b(t)$, and 
\begin{align}
    \mathcal{A}(\omega) = & (\lambda_a+\lambda)(j\omega) + (k_a+k) - m\omega^2, \\
    \mathcal{B}(\omega) = & \lambda_a(j\omega) +  k_a,                          \\
    \mathcal{C}(\omega) = & \lambda_a(j\omega) +  k_a - m_a\omega^2.            
\end{align}
Thus, we can obtain the following conditionally background variances
\begin{align}
    \sigma^2_{x_b} = \int^\infty_0 S_{x_bx_b}(\omega) \, d\omega, \quad                  
    \sigma^2_{\dot{x}_b} = \int^\infty_0 S_{\dot{x}_b\dot{x}_b}(\omega)\, d\omega, \quad 
    \sigma^2_{\ddot{x}_b} = \int^\infty_0 S_{\ddot{x}_b\ddot{x}_b}(\omega)\, d\omega,    \\
    \sigma^2_{y_b} = \int^\infty_0 S_{v_bv_b}(\omega)\, d\omega, \quad                   
    \sigma^2_{\dot{y}_b} = \int^\infty_0 S_{\dot{y}_b\dot{y}_b}(\omega)\, d\omega, \quad 
    \sigma^2_{\ddot{y}_b} = \int^\infty_0 S_{\ddot{y}_b\ddot{y}_b}(\omega)\, d\omega.    
\end{align}

\printbibliography 

\end{document}